\documentclass[aps,nofootinbib,amsmath,prc,showpacs,groupedaddress,12pt,superscriptaddress,notitlepage]{revtex4-1}

\usepackage{amsfonts}
\usepackage{amsmath}
\usepackage{amssymb}
\usepackage{epsfig}
\usepackage{latexsym}
\usepackage{paralist}
\usepackage{fancyhdr}
\usepackage{graphicx}
\usepackage{float}
\usepackage{slashed}
\usepackage{xcolor}
\usepackage{soul}
\usepackage{dcolumn}
\usepackage{physics}
\usepackage{amsmath}
\usepackage{hyperref}

\def\bea{\begin{eqnarray}}
\def\eea{\end{eqnarray}}
\def\be{\begin{equation}}
\def\ee{\end{equation}}
\def\ba{\begin{array}}
\def\ea{\end{array}}

\def\Tr{\mbox{Tr}}

\def\ep{\epsilon}

\def\beq{\begin{eqnarray}}    %%%    begequation/eqnarray
\def\eeq{\end{eqnarray}}       %%%    endequation/eqnarray
\def\D{\Delta_g}       %%% Imaginary
\def\al{\alpha}
\def\be{\beta}

\def\ga{\gamma}

\def\ep{\epsilon}

\def\la{\lambda}

\def\pa{\partial}

\def\si{\sigma}

\def\Ga{\Gamma}

\def\xibar{\left(\xi-\frac{1}{6} \right)}

\begin{document}
%%%%%%%%%%%%%%%%%%%%%%%%%%%%

\title{Form factors and decoupling of matter fields\\ in four-dimensional gravity}

\author{Sebasti{\'a}n~A.~Franchino-Vi{\~n}as}
\email{sa.franchino@uni-jena.de}
\affiliation{Theoretisch-Physikalisches Institut,
Friedrich-Schiller-Universit\"{a}t Jena, Max-Wien-Platz 1,
07743 Jena, Germany}

\author{Tib{\'e}rio~de~Paula~Netto}
\email{tiberiop@fisica.ufjf.br}
\affiliation{Departamento de F{\'i}sica, ICE, Universidade Federal de
Juiz de Fora, Juiz de Fora, 36036-100, Minas Gerais, Brazil}

\author{Ilya~L.~Shapiro}
\email{shapiro@fisica.ufjf.br}
\affiliation{Departamento de F{\'i}sica, ICE,
Universidade Federal de Juiz  de  Fora, Juiz de Fora,
36036-100, Minas Gerais, Brazil}
\affiliation{Tomsk State Pedagogical University, Tomsk,
634041, Russia}
\affiliation{Tomsk State University, Tomsk, 634050, Russia}

\author{Omar~Zanusso}
\email{omar.zanusso@uni-jena.de}
\affiliation{Theoretisch-Physikalisches Institut,
Friedrich-Schiller-Universit\"{a}t Jena, Max-Wien-Platz 1,
07743 Jena, Germany}

%%%%%%%%%%%%%%%
\begin{abstract}
%%%%%%%%%%%%%%%
\noindent
We extend previous calculations of the non-local form factors
of semiclassical gravity in $4D$ to include the Einstein-Hilbert term.
The quantized fields are massive
scalar, fermion and vector fields. The non-local form factor in this
case can be seen as the sum of a power series of total
derivatives, but it enables us to derive the beta function of Newton's constant
and formally evaluate the decoupling law in the new sector, which turns
out to be the standard quadratic one.
%%%%%%%%%%%%%%%
\end{abstract}
%%%%%%%%%%%%%%%

\maketitle

%%%%%%%%%%%%%%%%%%%%%%%%%%%%%%%%
\section{Introduction}\label{sect:introduction}
%%%%%%%%%%%%%%%%%%%%%%%%%%%%%%%%

The derivation of non-local form factors in the semiclassical theory
of massive matter fields on a classical curved background has several
interesting applications. The calculation in the higher derivative
vacuum sector \cite{apco,fervi} (see also \cite{Codello:2012kq})
supports the idea of the gravitational decoupling which is relevant for the
graceful exit from the general version of anomaly induced inflation
\cite{susykey,Shocom,asta}. Indeed, this mechanism is not sufficient
for deriving the Starobinsky inflation \cite{star,star83} from
quantum corrections, but one can hope that more detailed study of
the gravitational decoupling may be useful for constructing the
corresponding field theoretical model \cite{StabInstab}.

An important application of the effective approach to quantum field 
theory in curved spacetime is the possible running of cosmological
and Newton's constants at low energies, such as the typical energy
scale in the late cosmology (which we shall call IR). If such a 
running takes place, there could be measurable implications in both 
cosmology (see e.g.\ \cite{CC-Gruni}) and astrophysics (see, e.g., 
\cite{RotCurves}). Unfortunately, from the quantum field theory
side, there is no way to consistently calculate such a running. The
reason is that the existing methods of quantum calculations in curved
space are essentially based on the expansion of all quantities around
the flat space-time. For instance, the normal coordinate expansion
and Schwinger-DeWitt technique are based on the expansions into a power
series in the curvature tensor and its covariant derivatives. Such an
expansion is not sufficient to establish the physical running of the
cosmological and Newton's constants. An observation of such a
running  requires at least the  expansion around  space-times of
constant nonzero curvature \cite{DCCrun}, which is not available,
except some special cases \cite{Verd}, which are not sufficient to
observe the decoupling. In the case when a variation with respect to the scale of
the cosmological and Newton's constants does not take place, there
would be a discrepancy between the well established running of these
constants in the Minimal Subtraction ($\overline{\rm MS}$) renormalization scheme
\cite{nelpan82,buch84} (see \cite{book} for an introduction) and
the absence of the non-local form factors for the corresponding terms
in the effective action.

The reason why there are no non-local form factors in the
zero and second-derivative sectors of the gravitational action
can be easily seen from the comparison with the fourth-derivative
terms \cite{apco}. The non-local form factors can emerge in the
square of the Weyl tensor
$C_{\al\be\rho\si}\,k_1\big(\textstyle{\frac{\Box}{m^2}}\big)
C^{\al\be\rho\si}$, or
in the square of the scalar curvature
$R\,k_2\big(\textstyle{\frac{\Box}{m^2}}\big)R$.
At the same time it is unclear how to introduce such a form factor 
for the cosmological constant, because the d'Alembert operator 
acting on a constant gives zero. Furthermore, if a non-local form 
factor is inserted into the Einstein-Hilbert action, a function of 
$\Box$ acting on $R$  is equivalent to a sum of the series of surface 
terms. The simplest solution which was  proposed in \cite{apco} was 
to replace the cosmological constant by the non-local expressions
\beq
R_{\al\be}\,\frac{1}{\Box^2}\,R^{\al\be}
\qquad
\mbox{and}
\qquad
R\,\frac{1}{\Box^2}\,R,
\label{CCreplace}
\eeq
which have the same global scaling as a constant. Similar 
replacement can be done for the Einstein-Hilbert Lagrangian 
by using the terms
\beq
R_{\al\be}\,\frac{1}{\Box}\,R_{\al\be}
\qquad
\mbox{and}
\qquad
R\,\frac{1}{\Box}\,R \,.
\label{EHreplace}
\eeq
The problem with this approach is that the semiclassical form factors
can not be derived for the terms \eqref{CCreplace} and \eqref{EHreplace}
within the existing field theoretical methods. Thus, the interesting
cosmological applications of the models based on \eqref{CCreplace} and
\eqref{EHreplace} which were considered in \cite{Maggiore} are as
phenomenological as the non-covariant running which is considered
in \cite{CC-Gruni,DCCrun}, and the unique advantage, from the
conceptual point of view, is that those are covariant expressions,
which are easier to work with. In fact these structures are becoming increasingly
of interest even in the context of quantum gravity, in which they might play the role
of template to reconstruct the effective action \cite{Knorr:2018kog}.

Recently an alternative approach to the physical running of the
inverse Newton's constant has been initiated in \cite{Ribeiro:2018pyo}
which is based on \cite{Codello:2012kq}.
The consideration was performed for the two dimensional ($2D$) case
and is related to some older works by Avramidi and collaborators \cite{Avramidi:2007zz,Avramidi:1997hy}.
The idea is to derive the non-local form factors for the Einstein-Hilbert
term, regardless of the fact that the corresponding structures will be
total derivatives.\footnote{Similar structures have already been explored in the quantum gravity literature \cite{Hamber:2010an,Hamber:2011kc}.}
There is a serious justification of this approach,
but we postpone this part of the discussions for the last section. In what
follows we generalize the calculations of \cite{Ribeiro:2018pyo} to
four dimensions ($4D$) and perform full consideration of the
non-local terms. For the sake of completeness we checked all the
non-local contributions for higher derivative terms, which are well
known from \cite{apco,fervi} and \cite{Codello:2012kq}. One of
the reasons for this is the detailed discussion of the distinctions and
similarities between the form factors  for $R$, $\Box R$ and $R^2$
terms. As we know from previous work (see, e.g., the discussion in
\cite{anom2003} with a special emphasize to the role of non-local
form factors in massive semiclassical theory), the renormalization of
the surface terms results in the finite non-surface contributions, and
the explicit form of the non-local surface terms derived here makes
our understanding of this relation more detailed.

The outline of the paper is as follows. In Sec.~\ref{sect:non-local-effective-action} we discuss the structure
of the effective action and its renormalization, and construct
the necessary equations to observe the gravitational version of the
Applequist-Carazzone theorem \cite{AC}
for the Newton constant in $4D$
curved space. In Secs. \ref{sect:scalar}, \ref{sect:dirac} and
\ref{sect:proca} we give explicit formulas for nonminimally coupled
scalars, Dirac spinors and Proca fields respectively. Finally, in Sec.~\ref{sect:conclusions}
we draw our conclusions, present a general
analysis of the results and comment on possible physical
interpretations and the prospects of further developments.
The two Appendices are included to clarify further the main text,
namely in Appendix \ref{sect:heat-kernel} we briefly present the heat
kernel method which is used for the computations, while in Appendix
\ref{sect:uv-structure} we survey the ultraviolet structure of the
effective action and its physical implications.

%%%%%%%%%%%%%%%%%%%%%%%%%%%%%%%%%%%%%
%%%%%%%%%%%%%%%%%%%%%%%%%%%%%%%%%%%%%
\section{Nonlocal effective action}\label{sect:non-local-effective-action}
%%%%%%%%%%%%%%%%%%%%%%%%%%%%%%%%%%%%%
%%%%%%%%%%%%%%%%%%%%%%%%%%%%%%%%%%%%%

We are interested in the contribution to the vacuum effective action
of a set of free massive matter fields which includes $n_{\rm s}$
nonminimally coupled scalars, $n_{\rm f}$ Dirac fermions and
$n_{\rm p}$ Proca fields. The integration of the free matter fields
fluctuations on curved background leads to the expression
\beq
\Ga[g] &=& n_{\rm s} \Gamma_{\rm s}[g]
+ n_{\rm f} \Gamma_{\rm f}[g] + n_{\rm p} \Gamma_{\rm p}[g]\,,
\eeq
in which $\Gamma_{\rm s}[g]$, $\Gamma_{\rm f}[g]$ and
$\Gamma_{\rm p}[g]$ denote the individual contributions for a
single field of each matter specie. The individual contributions
are\footnote{Starting from this section we assume the Wick rotation
and all notations are Euclidean. The positively defined Laplacian
operator $\D$ is defined in Appendix A and
$R_{\mu\nu}=\pa_\la \Ga^\la_{\mu\nu}+\dots$. At the same time
in all physical discussions we use pseudo-Euclidean notations.}
\begin{equation}
 \begin{split}
 &
 \Gamma_{\rm s}[g]
 = \frac{1}{2} \Tr_{\rm s} \ln \left( \D + \xi R +m_{\rm s}^2\right) ,
\\
 & %% \qquad
 \Gamma_{\rm f}[g] = -\Tr_{\rm f} \ln \left(\slashed D+m_{\rm f}\right) ,
 \\
 &
 \Gamma_{\rm p}[g] = \frac{1}{2}\Tr_{\rm v} \ln\left( \delta_\mu^\nu \D
 +\nabla_\mu\nabla^\nu + R_\mu{}^\nu + \delta_\mu^\nu m_{\rm v}^2\right) ,
 \end{split}
\end{equation}
in which each trace is taken over the appropriate degrees of freedom 
and $\D$ is defined as positive in Euclidean space. A little work is needed 
to cast all functional traces in the same form. Squaring the Dirac operator 
we arrive at the expression
\begin{equation}
 \begin{split}
 \Gamma_{\rm f}[g] &= -\frac{1}{2} \Tr_{\rm f} 
 \ln \left( \D + \frac{R}{4} + m_{\rm f}^2\right)\,.
 \end{split}
\end{equation}
When dealing with the Proca operator we need to take care of the 
longitudinal
modes, which can be done in at least two equivalent ways \cite{bavi85,BuGui}
and results in
\begin{equation}
 \begin{split}
 \Gamma_{\rm p}[g] &= \frac{1}{2} \Tr_{\rm v} \ln\left(\delta_\mu^\nu \D+R_\mu{}^\nu+ \delta_\mu^\nu m_{\rm v}^2\right)
 - \frac{1}{2} \Tr_{\rm s} \ln\left(\D +m_{\rm v}^2\right)\,.
 \end{split}
\end{equation}
Now each trace acts on the logarithm of an operator of Laplace-type
\beq
 \Ga[g] &=& \frac{1}{2} \Tr \ln \left(\D +E +m^2\right)
\eeq
for an appropriate endomorphism $E$ acting on the field's bundle. A standard way to compute traces of Laplace-type operators
is to use the heat kernel. We can represent the above trace as an integral over the heat kernel proper time $s$,
\begin{equation}
 \begin{split} \label{eq:effective-action-divergent}
 \Ga[g] &= -\frac{1}{2} \tr \int_0^\infty \frac{{\rm d}s}{s} \int{\rm d}^4x\sqrt{g} ~ {\rm e}^{-sm^2} {\cal H}(s;x,x)\,,
 \end{split}
\end{equation}
in which we have also separated the original trace into an integration
over spacetime and a trace over the internal indices, and introduced the local
heat kernel ${\cal H}(s;x,x')$ (see Appendix \ref{sect:heat-kernel} for a brief
explanation regarding the heat kernel technique).

The effective action \eqref{eq:effective-action-divergent} has ultraviolet
divergencies, and a simple way to regulate them is through dimensional
regularization \cite{bro-cass}.
For this purpose we continue
the leading power $s^{-\frac{d}{2}}$ of the heat kernel
% integral from four
to a generic number $d$ of dimensions,
%
% ${\rm d}^4x \to {\rm d}^dx$
% \begin{equation}
%  \begin{split}
%  & {\rm d}^4x ~ \to ~ {\rm d}^dx ~ \mu^{d-4} = \mu^{-\epsilon} ~ {\rm d}^dx \,,
%  \end{split}
% \end{equation}
% in which we have also introduced
and introduce both a reference scale $\mu$ to preserve the mass dimension of all quantities
when leaving four dimensions and a small parameter $\epsilon=4-d$.
The result of this substitution is the regularized effective action
\begin{equation}
 \begin{split} \label{eq:effective-action-regularized}
 \Ga[g] &= -\frac{\mu^{\epsilon}}{2} 
 \tr \int_0^\infty \frac{{\rm d}s}{s} 
 \int{\rm d}^4x \sqrt{g}~ {\rm e}^{-sm^2} {\cal H}(s;x,x).
 \end{split}
\end{equation}
Since all fields are massive the above effective action has no 
infrared divergences,  thanks to the exponential damping factor 
caused by the mass for large values of $s$. However, there are 
ultraviolet divergences which appear as inverse powers of 
$\epsilon$ and require renormalization. We follow the 
standard practice of subtracting poles of the parameter 
$\bar{\epsilon}$, which is defined as
\begin{equation}
 \begin{split}
 \frac{1}{\bar{\epsilon}}
 &
 = \frac{1}{\ep} + \frac{1}{2}\ln \left(\frac{4\pi\mu^2}{m^2}\right)
- \frac{\ga}{2}
\end{split}
\end{equation}
(here $\ga$ is the Euler's constant),
instead of simply subtracting $\epsilon$ poles, exploiting the freedom of the
choice of renormalization scheme.

In the process of regularization and renormalization it is often
convenient to deal with dimensionless quantities. Keeping in mind
that at the moment the energy scales at our disposal are the Laplacian
$\Delta_g$ and the mass $m^2$, we find convenient to introduce the
following dimensionless operators
\begin{equation}
\label{eq:dimensionless-operators}
 z = \frac{\D}{m^2}\,,
 \qquad
 a = \sqrt{\frac{4z}{4+z}}\,,
 \qquad
Y = 1-\frac{1}{a} \ln\left|{\frac{1+a/2}{1-a/2}}\right|\,.
\end{equation}
With the above definitions we have all the ingredients to discuss the
form that the effective action can take up to the second order in a
curvature expansion. We have that to this order the most general form
can be narrowed down to the sum of a local and a non-local part
\begin{equation}
 \begin{split} \label{eq:effective-action-full}
\Ga[g] &=
\Ga_{\rm loc}[g]
 + \frac{m^2}{2(4\pi)^2}\int {\rm d}^4 x \sqrt{g}\,
 B(z) R
 \\
 &
 + \frac{1}{2(4\pi)^2}\int {\rm d}^4 x \sqrt{g}\, \Bigl\{
 C^{\mu\nu\alpha\beta} \, C_1(z) \, C_{\mu\nu\alpha\beta}
 + R \, C_2(z) \, R
 \Bigr\}\,,
 \end{split}
\end{equation}
in which $C_{\mu\nu\rho\theta}$ is the four dimensional Weyl tensor.
Since the divergences are local expressions,  all dimensional poles are
contained in the local part of effective action $\Ga_{\rm loc}[g]$.
The renormalization can be performed through the introduction of
appropriate counter terms and generically results in a renormalized
action of the form
\begin{equation}
 \begin{split}
 \label{eq:effective-action-local-renormalized}
 S_{\rm ren}[g]
 &= \int {\rm d}^4 x \sqrt{g}\,\left\{b_0 + b_1 R + a_1 C^2 +a_2 {\cal E}_4
 +a_3 \Box R + a_4 R^2\right\}\,,
 \end{split}
\end{equation}
in which $\,{\cal E}_4$ is the operator associated to the Euler's
characteristic, which is the Gauss-Bonnet topological term in $d=4$.
The renormalized action features the couplings that have to be
experimentally determined in order for the theory to be predictive.
The couplings include the cosmological constant $\Lambda$ and
the Newton's constant $G$ through the relations $b_0=2\Lambda G^{-1}$
and $b_1=-G^{-1}$.
The minimal subtraction ($\overline{\rm MS}$) procedure induces a
running of all the couplings which is encoded in beta functions that we
denote as $\beta^{\overline{\rm MS}}_g$ in which $g$ is any of the
couplings of \eqref{eq:effective-action-local-renormalized}. In what
follows we formulate the beta functions for the parameters $b_0$ and
$b_1$, instead of $\Lambda$ and $G$.

The minimal subtraction scheme - based one-loop  renormalization
group flow induced by the beta functions
of the couplings of \eqref{eq:effective-action-local-renormalized}
has been known for a long time for all the field types listed in this
section. In this work we concentrate instead on the non-local
contributions of the effective action. In \eqref{eq:effective-action-full}
we have introduced three new covariant functions $B(z)$, $C_1(z)$
and $C_2(z)$ of the rescaled Laplacian $z$. These functions are known
as form factors of the effective action and represent a true physical
prediction which comes from the formalism: in fact one can imagine
to pick a specific observable -- either from cosmology or from particle
physics -- and compute it in terms of the form factors themselves
\cite{Donoghue:2017pgk}. A simple way to understand the physical
consequences of the effective action, which is  related to the general
concept of renormalization group, is to use them to construct new
non-local beta functions which are  sensitive to the presence of the
mass scale $m^2$.

Let us first recall that the non-local form factors of the heat kernel of Appendix \ref{sect:heat-kernel},
and consequently the non-local contributions to the effective action \eqref{eq:effective-action-full},
are obtained for asymptotically flat Euclidean spacetimes in which curvatures are small
(schematically $\left|\nabla^2 {\cal R}\right| \gg \left|{\cal R}^2\right|$ for any curvature tensor ${\cal R}$) \cite{bavi90}.
In practice, the asymptotic flatness offers a special reference frame which can be used to construct meaningful
Fourier transformations and in which the expansion in curvatures can be related to
the expansion in fluctuations of the metric. In fact, this is precisely the frame in which the form factors are computed
in \cite{apco,fervi,Codello:2012kq}, even though the final expressions are always presented in a manifestly covariant form.
In short, this implies that the Laplace operator $\Delta_g$ is in one-to-one correspondence with
the square $q^2$ of a momentum $q_\mu$ of the asymptotic frame upon Fourier transformation.
This representation is especially useful for the renormalization group applications,
where one has to take derivatives with respect to the scale parameter.

The straightforward way to derive the beta functions is to subtract
the divergences at the scale $q=\left|q_\mu\right|$.
%defined as the argument of the Fourier transform of the form factors.
For convenience, let us define the
dimensionless scale $\hat{q}= q/m$ which is simply $q$ in units of the mass; by
definition after the Fourier transform $\hat{q}$ is related to
$z$ as $\hat{q}^2=z$ and the renormalization group flow is
parametrized by
\begin{equation}
 \begin{split}
 q\frac{\partial}{\partial q}
 ~ = ~ \hat{q} \frac{\partial}{\partial \hat{q}}
  ~ = ~ 2 z \frac{\partial}{\partial z}\,.
 \end{split}
\end{equation}

Let us begin by discussing the renormalization group flow of the
terms that are quadratic in the curvatures which has been studied
in detail in \cite{apco,fervi}. A simple inspection suggests the
non-local generalization of the beta functions of $a_1$
\begin{equation}
 \begin{split}
 \beta_{a_1}
 ~=~
 2 z \,\frac{\partial}{\partial z}\,
 \left[\frac{1}{2(4\pi)^2} C_1(z)\right]
 ~ = ~
 \frac{z}{(4\pi)^2} C'_1(z)\,,
\end{split}
\end{equation}
in which we indicate the derivative with a prime. The same can
be done for the coupling $a_4$
\begin{equation}
 \begin{split}
 \beta_{a_4} ~= ~ \frac{z}{(4\pi)^2} C'_2(z)\,.
 \end{split}
\end{equation}
In practice the form factors $C_1(z)$ and $C_2(z)$ play the role of
non-local scale-dependent generalizations of the couplings. Since
our heat kernel methods work on spaces that are asymptotically flat,
we do not have enough information to compute the running of the
topological term in this context (although it is still possible to
complement this result with standard Seeley-DeWitt methods).

Now let us turn our attention to the couplings of the terms that are
linear in the curvature $R$. On the one hand we have that the
renormalized action features two couplings -- $b_1$ and $a_3$ --
but on the other hand there is only a single form factor $B(z)$
acting on $R$ in \eqref{eq:effective-action-full}. Naively we are
tempted to define a master beta function
\begin{equation}\label{eq:def-psi}
 \begin{split}
 \Psi ~ = ~ \frac{1}{(4\pi)^2}\, z\,\partial_z \left[\frac{B(z)}{z}\right],
 \end{split}
\end{equation}
which we denoted with a new symbol to avoid confusion. The running
function $\Psi$ is defined to take into account that in
\eqref{eq:effective-action-full} we are measuring a dimensionfull
quantity -- the coefficient of $R$ -- in units of $m^2$ while instead
our rescaling should be done in units of $q^2$, hence the quotient
with $z=q^2/m^2$ that restores the right units. While we will find
useful to study this object later on, at this stage it is not clear if its
renormalization group flow should be associated to the coupling
$b_1$ or to $a_3$.
Returning to \eqref{eq:effective-action-local-renormalized} it is easy
to see that if $m^2\gg q^2$ the operator $R$ will dominate
over the operator $\Box R$, and conversely if $m^2\ll q^2$ the operator
$\Box R$ will dominate over the operator $R$. This implies that in the
high energy limit $z\gg 1$ the function $\Psi$ should encode information
of $\beta_{a_3}$, while in the opposite limit $z \sim 0$ the function
$\Psi$ should encode information of $\beta_{b_1}$. This property is
discussed in more detail later. However, $\beta_{b_1}$ and $\beta_{a_3}$
have well known ultraviolet limits which we would like to preserve,
associated to $\overline{\rm MS}$ as we will also see later.
We find that the best solution is to define the following beta functions
\begin{equation}
 \begin{split}
 \label{eq:definitions-running}
 \beta_{a_3} ~ = ~ -\frac{1}{(4\pi)^2} \,z\,\partial_z
 \left[\frac{B(z)-B(0)}{z}\right],
 \qquad
 \beta_{b_1} ~ = ~ \frac{m^2}{(4\pi)^2} \,z\,
 \partial_z\big[B(z)-B_\infty(z)\big].
 \end{split}
\end{equation}
The first equation is implied by the comparison with
\eqref{eq:effective-action-local-renormalized} and
% essentially
% provides the definition $\beta_{a_3}=-\Psi$.
includes the removal of the constant part that should be attributed to $b_1$.
In the second equation
we subtract the dominating $\Box R$ effect from the running of
$B(z)$ in the form of $B_\infty(z)$
which is the leading logarithmic asymptotic behavior for
$z\simeq\infty$ of $B(z)$ itself. The leftover terms of the
subtraction is thus identified with the running of the operator
$R$ and hence the coupling $b_1$. In the practical computations
instead of subtracting the leading logarithm  at infinity,
we will subtract instead the combination
\begin{equation}
 \begin{split}
  a(1-Y)\simeq \ln (z) \,,
 \end{split}
\end{equation}
which is shown to be valid for $z\gg 1$ using the definitions
\eqref{eq:dimensionless-operators}. General features of the definitions
\eqref{eq:definitions-running} and their ultraviolet properties are
discussed in more detail in Appendix \ref{sect:uv-structure}.

In the next section we present explicit results for the form factor and
the beta function for the Einstein-Hilbert term. The full set of the
form factors and the expressions for all the non-local beta functions
in the fourth-derivatives sector, and also the corresponding
$\overline{\rm MS}$ beta functions can be found in the
papers \cite{fervi,Codello:2012kq}.
All results will be collected in a mini review to appear shortly
\cite{minireview}.
However, there are still some general
properties that we can discuss here in anticipation. For all the
couplings and all the beta functions we can show that there are
sensible ultraviolet $z\gg 1$ and infrared $z \sim 0$ limits.
Each beta function satisfies the additional property
\begin{equation}
 \begin{split}
\beta_g
&= \beta^{\overline{\rm MS}}_g
+ {\cal O}\left(\frac{m^2}{q^2}\right) \qquad {\rm for }
\qquad
q^2\gg m^2,
 \end{split}
\end{equation}
where $g$ is any of the couplings.
Furthermore, all the renormalization group running the subject to
the effect of decoupling towards the infrared, meaning that when
$q^2$ goes below the $m^2$ threshold fluctuations stop propagating
and have no effect on the quantum physics anymore. We have that
\begin{equation}
 \begin{split}
\beta_g
&=  {\cal O}\left(\frac{q^2}{m^2}\right) \qquad {\rm for }\qquad q^2 \ll m^2,
 \end{split}
\end{equation}
which is the practical evidence of the Applequist-Carazzone theorem
in four dimensional curved space.

Finally, it is interesting to observe the practical implications of
the discussion on the function $\Psi(z)$. As argued above, the
limits $m^2\ll q^2$ and $m^2\gg q^2$ should see the operators
$\Box R$ and $R$ dominating the running $\Psi(z)$ respectively.
For all the matter types that we consider we have the following two
limits
\begin{equation}
 \begin{split}
  \Psi &=  \begin{cases}
- \beta^{\overline{\rm MS}}_{a_3}
& \qquad {\rm for} \quad q^2 \gg m^2
 \\
 \frac{m^2}{q^2} ~ \beta^{\overline{\rm MS}}_{b_1}
 & \qquad {\rm for} \quad q^2 \ll m^2
 \end{cases}
 \end{split}
\end{equation}
which reflect the previous consideration. Notice that while the
ultraviolet limit can be straightforwardly proven on the basis of
the definitions of $\beta^{\overline{\rm MS}}_{a_3}$ and $\Psi$,
the infrared limit is much less trivial. Notice also that the infrared
limit does not sharply decouple, because it grows with the square
of the mass, but this is to be expected since we are measuring a
massive quantity in units of $q$ for $q\to 0$. To get rid of the
divergence it is sufficient to switch to measuring the same quantity
in units of $m$ in the infrared.

%%%%%%%%%%%%%%%%%%%%%%%%%%%%%
%%%%%%%%%%%%%%%%%%%%%%%%%%%%%
\section{Nonminimally coupled scalar field}\label{sect:scalar}
%%%%%%%%%%%%%%%%%%%%%%%%%%%%%

\begin{figure}
 \includegraphics[height=4.5cm]{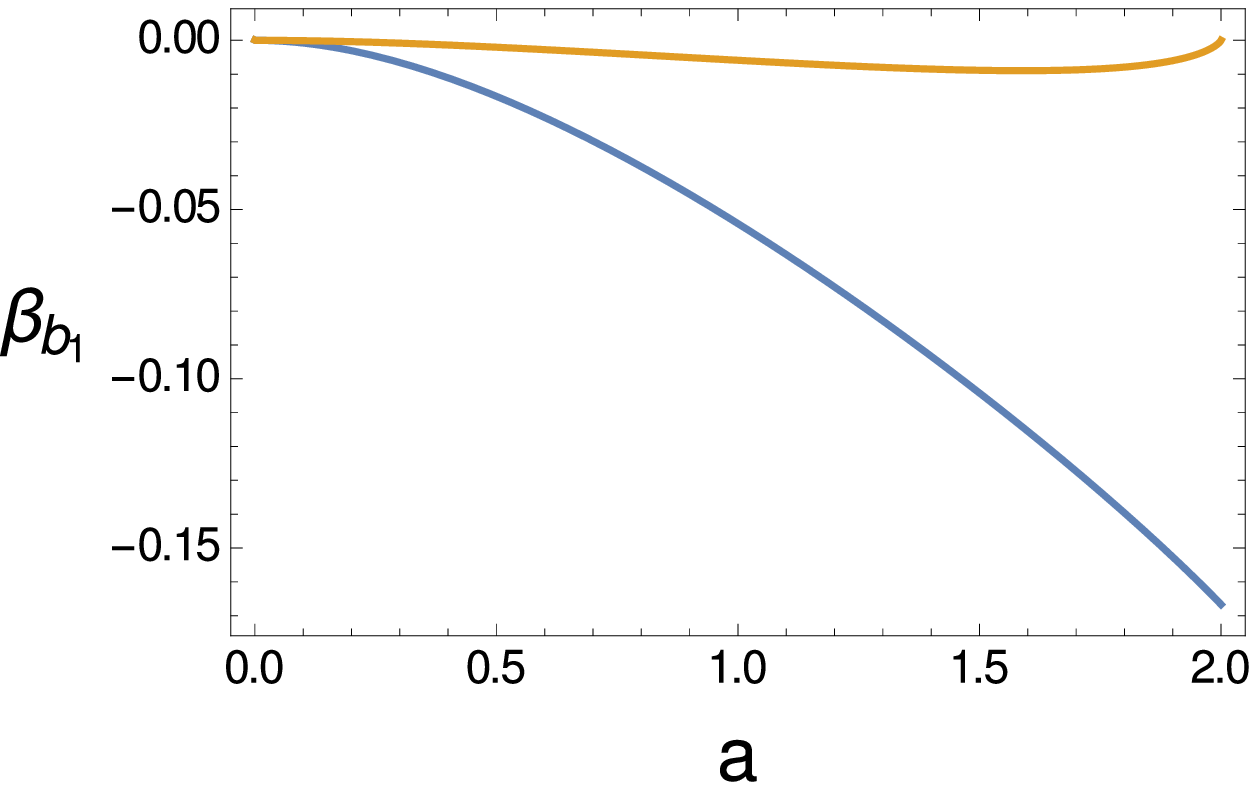}\qquad \includegraphics[height=4.5cm]{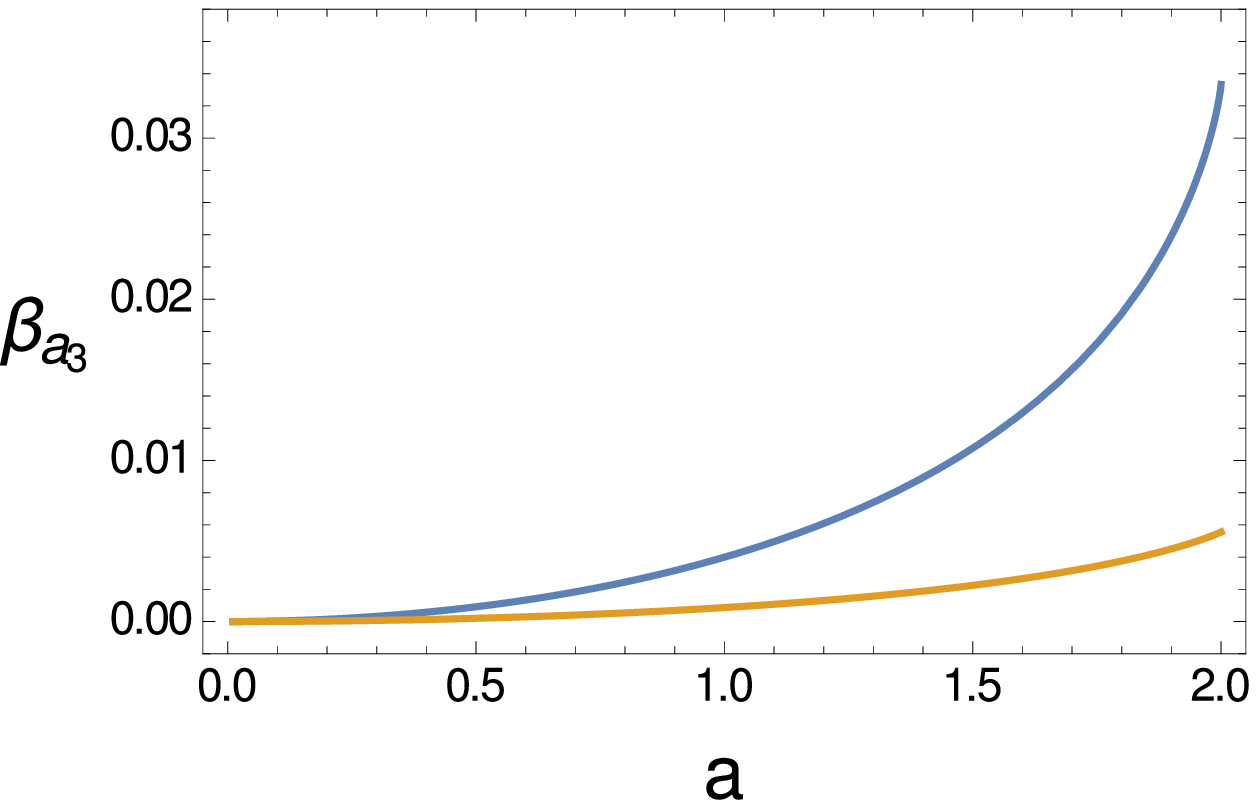}
 \caption{Plots of the beta functions $\beta_{b_1}$ and $\beta_{a_3}$ rescaled by a factor $(4\pi)^2$
 that are induced by a single scalar field for the values $\xi=0$ (blue) and $\xi=\frac{1}{6}$ (yellow)
 as a function of the variable $a$ defined in \eqref{eq:dimensionless-operators}. The plot ranges from the IR at $a=0$ ($q^2\ll m^2$) to the UV at $a=2$ ($q^2\gg m^2$).
 The effects of the Applequist-Carazzone theorem are seen on the left where the beta functions become zero. The beta function $\beta_{b_1}$ for the special conformal value $\xi=\frac{1}{6}$
 is zero also in the UV.}
 \label{figure:plots-scalar}
\end{figure}

The effective action of the nonminimally coupled scalar field
can be obtained specifying the endomorphism $E=\xi R$
in the non-local heat kernel expansion and then performing
the integration in $s$ \cite{Martini:2018ska}.
We find the local contributions of the regularized action to be
\begin{equation}
 \begin{split}
 \Ga_{\rm loc}[g] &=
 \frac{1}{2(4\pi)^2} \int {\rm d}^4 x\sqrt{g} \, \Bigl\{
 -m^4\Bigl(\frac{1}{\bar{\epsilon}}+\frac{3}{4}\Bigr)
 - 2m^2\Bigl( \xi-\frac{1}{6} \Bigr)\frac{1}{\bar{\epsilon}} R
  \\ & \qquad\qquad
 +\frac{1}{3} \Bigl( \xi-\frac{1}{5} \Bigr)\frac{1}{\bar{\epsilon}} \Box R
 -\frac{1}{60\bar{\epsilon}} C_{\mu\nu\rho\theta} C^{\mu\nu\rho\theta}
 -\Bigl( \xi-\frac{1}{6} \Bigr)^2 \frac{1}{\bar{\epsilon}} R^2
 \Bigr\}\,.
 \end{split}
\end{equation}
The minimal subtraction of the divergences of local contributions
induces the following $\overline{\rm MS}$ beta functions
for the terms with up to one curvature
\begin{equation}
\begin{array}{lll}
 \beta_{b_0}^{\overline{\rm MS}} = \frac{1}{(4\pi)^2} \frac{m^4}{2} \,,
 &  \qquad
 \beta_{b_1}^{\overline{\rm MS}} = \frac{1}{(4\pi)^2} m^2 \xibar \,,
 & \qquad
 \beta_{a_3}^{\overline{\rm MS}} = - \frac{1}{(4\pi)^2} \frac{1}{6} \left(\xi-\frac{1}{5}\right)\,.
\end{array}
\end{equation}
The non-local part of the effective action includes the following form factor
\begin{equation}
 \begin{split}
 \frac{B(z)}{z} &=
-\frac{4 Y}{15 a^4}+\frac{Y}{9 a^2}-\frac{1}{45 a^2}
+\frac{4}{675}+\xibar \left(-\frac{4 Y}{3 a^2}-\frac{1}{a^2}
+\frac{5}{36}\right)\,,
 \end{split}
\end{equation}
while $C_1(z)$ and $C_2(z)$ confirm the results reported in
\cite{apco}. Using our definitions the non-local beta functions
of the couplings associated to the curvature $R$ are
\beq
\beta_{b_1}
&=&
\frac{z}{(4\pi)^2}\Bigl\{
 \frac{2 Y}{5 a^4}-\frac{2 Y}{9 a^2}
 + \frac{1}{30 a^2}-\frac{aY}{180}
 +\frac{a}{120}+\frac{Y}{24}-\frac{1}{40}
  \nonumber
  \\
  &+&
  \xibar \left(\frac{2 Y}{3 a^2}+\frac{a Y}{6}-\frac{a}{4}
  -\frac{Y}{2}+\frac{1}{2}\right)
  \Bigr\}
\label{Gff}
\eeq
and
\beq
 \beta_{a_3} &=& \frac{1}{(4\pi)^2}
 \Bigl\{
 -\frac{2 Y}{3 a^4}
 +\frac{Y}{3 a^2}-\frac{1}{18 a^2}
 -\frac{Y}{24}+\frac{7}{360}
\nonumber
\\
&+&
\xibar
 \left(-\frac{2 Y}{a^2}+\frac{Y}{2}-\frac{1}{6}\right)
 \Bigr\}\,.
\label{a3ff}
\eeq
The Eqs.~\eqref{Gff} and \eqref{a3ff} provide all necessary
ingredients to study the Applequist-Carazzone theorem of both
parameters.
Plots of these beta functions are given in Fig.~\ref{figure:plots-scalar}.

As we have explained in the Introduction, the most interesting is
the decoupling theorem for the running of the Newton's constant
which is related to the inverse of $b_1=-G^{-1}$. The non-local
beta function of the couplings $b_1$ and $a_3$ in units of the
mass have the two limits
\begin{equation}
\begin{split}
\frac{\beta_{b_1}}{m^2} &=  \begin{cases}
\frac{1}{(4\pi)^2}\xibar + \frac{1}{(4\pi)^2}
\left\{\left(\frac{3}{5}-\xi\right)
-\xi \ln\left(\frac{q^2}{m^2}\right)\right\} \frac{m^2}{q^2}
+{\cal O}\left(\frac{m^2}{q^2}\right)^{2}
&
\quad {\rm for} \quad q^2 \gg m^2,
\\
\frac{1}{(4\pi)^2}\left(\frac{4}{9}\xi-\frac{77}{900}\right)
\frac{q^2}{m^2} +{\cal O}\left(\frac{q^2}{m^2}\right)^{\frac{3}{2}}
&
\quad {\rm for} \quad q^2 \ll m^2
\end{cases}%\end{cases}
 \end{split}
\end{equation}
and
\begin{equation}
\begin{split}
\beta_{a_3} &= \begin{cases}
-\frac{1}{6(4\pi)^2}\left(\xi-\frac{1}{5}\right)
+ \frac{1}{(4\pi)^2}  \left\{ \frac{5}{18}-2\xi
+ \xibar \ln\left(\frac{q^2}{m^2}\right)\right\} \frac{m^2}{q^2}
            +{\cal O}\left(\frac{m^2}{q^2}\right)^{2}
& \,\,\, {\rm for} \,\,\, q^2 \gg m^2,
\\
\frac{1}{(4\pi)^2} \frac{1}{840}
\left(3-14\xi\right)\frac{q^2}{m^2}
+{\cal O}\left(\frac{q^2}{m^2}\right)^2
& \,\,\, {\rm for} \,\,\, q^2 \ll m^2.
\end{cases}
 \end{split}
\end{equation}
The last expressions show standard quadratic decoupling in the IR
for both parameters,
exactly as in the usual QED situation \cite{AC} and as for the fourth
derivative non-surface gravitational terms \cite{apco,fervi}. In the high
energy limit (UV)
we meet the usual $\overline{\rm MS}$ beta function plus a small correction
to it.

%%%%%%%%%%%%%%%%%%%%%%%%%%%%%%%%%%%%%%%%%%%%%%%%%%%%%%%%%%%%%%%
%%%%%%%%%%%%%%%%%%%%%%%%%%%%%%%%%%%%%%%%%%%%%%%%%%%%%%%%%%%%%%%
\section{Dirac field}\label{sect:dirac}
%%%%%%%%%%%%%%%%%%%%%%%%%%%%%%%%%%%%%%%%%%%%%%%%%%%%%%%%%%%%%%%
%%%%%%%%%%%%%%%%%%%%%%%%%%%%%%%%%%%%%%%%%%%%%%%%%%%%%%%%%%%%%%%

The effective action of the minimally coupled Dirac fields requires the specification of the endomorphism $E=R/4$.
The final result turns out to be proportional to the dimension $d_\gamma$ of the Clifford algebra and hence to
the number of spinor components. We do not set $d_\gamma=4$, but choose instead to leave it arbitrary
so that the formulas can be generalized to other spinor species easily.
We find the local regularized action to be
\begin{equation}
 \begin{split}
 \Ga_{\rm loc}[g] &=\frac{d_\gamma}{2(4\pi)^2} \int {\rm d}^4 x\sqrt{g} \, \Bigl\{
 m^4\Bigl(\frac{1}{\bar{\epsilon}}+\frac{3}{4}\Bigr)
 +\frac{m^2}{6\bar{\epsilon}} R -\frac{1}{60\bar{\epsilon}} \Box R
 -\frac{1}{40\bar{\epsilon}} C_{\mu\nu\rho\theta} C^{\mu\nu\rho\theta}
 \Bigr\}\,.
 \end{split}
\end{equation}
The minimal subtraction of the $1/\bar{\epsilon}$ divergences induces the following $\overline{\rm MS}$ beta functions
\begin{equation}
\begin{array}{lll}
 \beta_{b_0}^{\overline{\rm MS}}
 = -\frac{d_\gamma}{(4\pi)^2} \frac{m^4}{2} \,,
 &  \qquad
 \beta_{b_1}^{\overline{\rm MS}}
 = -\frac{d_\gamma}{(4\pi)^2} \frac{m^2}{12} \,,
 & \qquad
 \beta_{a_3}^{\overline{\rm MS}}
 = \frac{d_\gamma}{(4\pi)^2} \frac{1}{120} \,.
\end{array}
\end{equation}
The non-local part of the effective action includes the following form factor
\begin{equation}
 \begin{split}
 \frac{B(z)}{z} &=
 d_\gamma\Bigl\{-\frac{7}{400}+\frac{19}{180a^2}+\frac{4Y}{15a^4} \Bigr\}\,,
 \end{split}
\end{equation}
while $C_1(z)$ and $C_2(z)$ agree with \cite{apco}.
The non-local beta functions are
\begin{equation}
 \begin{split}
 \beta_{b_1} &= \frac{d_\gamma z}{(4\pi)^2}\Bigl\{-\frac{2 Y}{5 a^4}+\frac{Y}{6 a^2}-\frac{1}{30 a^2}-\frac{a Y}{120}+\frac{a}{80}-\frac{1}{60} \Bigr\}\,,
  \\
  \beta_{a_3} &= \frac{d_\gamma}{(4\pi)^2} \Bigl\{ \frac{2 Y}{3 a^4}-\frac{Y}{6 a^2}+\frac{1}{18 a^2}-\frac{1}{180} \Bigr\}\,.
 \end{split}
\end{equation}
%\frac{2 Y}{3 a^4}-\frac{Y}{6 a^2}+\frac{1}{18 a^2}-\frac{1}{180}
%
Likewise the scalar case the non-local beta functions of $b_1$ and $a_3$
have the two limits
\begin{equation}
\begin{split}
\frac{\beta_{b_1}}{m^2}
&=  \begin{cases}
            -\frac{d_\gamma}{(4\pi)^2}\frac{1}{12}
            -\frac{d_\gamma}{(4\pi)^2}\left[\frac{7}{20}
            -\frac{1}{4}\ln\left(\frac{q^2}{m^2}\right)\right]\frac{m^2}{q^2} +{\cal O}\left(\frac{m^2}{q^2}\right)^{{2}}
& \qquad {\rm for} \quad q^2 \gg m^2\,;
\\
            -\frac{d_\gamma}{(4\pi)^2}\frac{23}{900} \frac{q^2}{m^2} +{\cal O}\left(\frac{q^2}{m^2}\right)^{\frac{3}{2}}
& \qquad {\rm for} \quad q^2 \ll m^2\,.
           \end{cases}\\
  \beta_{a_3} &= \begin{cases}
            \frac{d_\gamma}{(4\pi)^2} \frac{1}{120} + \frac{d_\gamma}{(4\pi)^2}\left\{\frac{2}{9}
            -\frac{1}{12}\ln\left(\frac{q^2}{m^2}\right)\right\}\frac{m^2}{q^2}
            +{\cal O}\left(\frac{m^2}{q^2}\right)^{{2}}
            & \qquad {\rm for} \quad q^2 \gg m^2 \,;
            \\
            \frac{d_\gamma}{(4\pi)^2} \frac{1}{1680} \frac{q^2}{m^2}
            +{\cal O}\left(\frac{q^2}{m^2}\right)^2
            & \qquad {\rm for} \quad q^2 \ll m^2\,.
           \end{cases}
 \end{split}
\end{equation}
Once again, there is a standard quadratic decoupling in the IR
for both parameters, while in the UV we find the $\overline{\rm MS}$
beta function and a sub-leading correction.

%%%%%%%%%%%%%%%%%%%%%%%%%%%%%
%%%%%%%%%%%%%%%%%%%%%%%%%%%%%
\section{Proca field}\label{sect:proca}
%%%%%%%%%%%%%%%%%%%%%%%%%%%%%
%%%%%%%%%%%%%%%%%%%%%%%%%%%%%

The minimally coupled Proca field could be understood as a
four-components vector field, but one of these components is subtracted
through a single scalar ghost, so it has effectively three degrees of
freedom in four dimensions. The local regularized action is
\begin{equation}
 \begin{split}
 \Ga_{\rm loc}[g] &=
 \frac{1}{2(4\pi)^2} \int {\rm d}^4 x\sqrt{g} \, \Bigl\{
 -m^4\Bigl(\frac{3}{\bar{\epsilon}}+\frac{9}{4}\Bigr)
 -\frac{m^2}{\bar{\epsilon}} R +\frac{2}{15\bar{\epsilon}} \Box R
 -\frac{13}{60\bar{\epsilon}} C_{\mu\nu\rho\theta} C^{\mu\nu\rho\theta}
 -\frac{1}{36} R^2
 \Bigr\}\,.
 \end{split}
\end{equation}
The minimal subtraction of the $1/\bar{\epsilon}$ poles induces the
following $\overline{\rm MS}$ beta functions
\begin{equation}
\begin{array}{lll}
 \beta_{b_0}^{\overline{\rm MS}} = \frac{1}{(4\pi)^2} \frac{3m^4}{2} \,,
 &  \qquad
 \beta_{b_1}^{\overline{\rm MS}} = \frac{1}{(4\pi)^2} \frac{m^2}{2} \,,
 & \qquad
 \beta_{a_3}^{\overline{\rm MS}} = -\frac{1}{(4\pi)^2} \frac{1}{15} \,.
\end{array}
\end{equation}
The non-local part of the effective action includes the following form factors
\begin{equation}
 \begin{split}
 \frac{B(z)}{z} &= \frac{157}{1800} -\frac{17}{30 a^2} -\frac{4 Y}{5 a^4}-\frac{Y}{3 a^2} \,,
 \end{split}
\end{equation}
and $C_1(z)$ and $C_2(z)$ reproduce \cite{apco}. The non-local beta functions are
\begin{equation}
 \begin{split}
 \beta_{b_1} &= \frac{z}{(4\pi)^2} \left\{ \frac{6 Y}{5 a^4}-\frac{Y}{3 a^2}+\frac{1}{10 a^2}+\frac{a Y}{15}-\frac{a}{10}-\frac{Y}{8}+\frac{7 }{40} \right\}
  \\
  \beta_{a_3} &= \frac{1}{(4\pi)^2} \Bigl\{ -\frac{2 Y}{a^4}-\frac{1}{6 a^2}+\frac{Y}{8}-\frac{1}{40} \Bigr\}\,.
 \end{split}
\end{equation}
% -\frac{2 Y}{a^4}-\frac{1}{6 a^2}+\frac{Y}{8}-\frac{1}{40}
%
The beta functions of $b_1$ and $a_3$ have the two limits
\begin{equation}
 \begin{split}
  \frac{\beta_{b_1}}{m^2} &=  \begin{cases}
            \frac{1}{(4\pi)^2}\frac{1}{2}+\frac{1}{(4\pi)^2}\left(\frac{4}{5}
            -\ln\left(\frac{q^2}{m^2}\right)\right)\frac{m^2}{q^2} +{\cal O}\left(\frac{m^2}{q^2}\right)^{2}
  & \qquad {\rm for} \quad q^2 \gg m^2\,;
            \\
            \frac{1}{(4\pi)^2}\frac{169}{900} \frac{q^2}{m^2}
            +{\cal O}\left(\frac{q^2}{m^2}\right)^{\frac{3}{2}}
  & \qquad {\rm for} \quad q^2 \ll m^2\,.
           \end{cases}\\
  \beta_{a_3} &= \begin{cases}
            -\frac{1}{(4\pi)^2} \frac{1}{15}
            - \frac{1}{(4\pi)^2}\left\{\frac{7}{6}
            -\frac{1}{2}\ln\left(\frac{q^2}{m^2}\right)\right\}\frac{m^2}{q^2}
            +{\cal O}\left(\frac{m^2}{q^2}\right)^{2}
            & \qquad {\rm for} \quad q^2 \gg m^2 \,;
            \\
            -\frac{1}{(4\pi)^2} \frac{1}{168} \frac{q^2}{m^2}
            +{\cal O}\left(\frac{q^2}{m^2}\right)^2 & \qquad {\rm for}
            \quad q^2 \ll m^2\,.
           \end{cases}
 \end{split}
\end{equation}
We can observe that for the Proca field there is the same quadratic
decoupling for both couplings, and the same $\overline{\rm MS}$
beta function plus a small correction in the UV.

%%%%%%%%%%%%%%%%%%%%%%%%%%%%%%%%%%%
%%%%%%%%%%%%%%%%%%%%%%%%%%%%%%%%%%%
\section{Conclusions}
\label{sect:conclusions}
%%%%%%%%%%%%%%%%%%%%%%%%%%%%%%%%%%%
%%%%%%%%%%%%%%%%%%%%%%%%%%%%%%%%%%%

We computed the covariant non-local form factors of the Euclidean
effective action of nonminimal scalars, Dirac spinors and Proca fields
up to the second order of the curvature expansion on asymptotically
flat space. The calculations were performed by means of heat kernel
method for the massive quantum fields and an arbitrary external
metric. We checked explicitly that the results for the fourth derivative
terms confirmed the
previous ones derived by \cite{apco,fervi,Codello:2012kq} which were
obtained by both Feynman diagrams and heat kernel method as presented
in the paper of Barvinsky and Vilkovisky \cite{bavi90}.  We used the
results for the effective action to find suitable beta functions
which arise from the subtraction of the divergences at a physical
momentum scale $q^2$. These beta functions are special because
they display two important limits: in the ultraviolet they reproduce
the universal results coming from the minimal subtraction of the
poles of dimensional regularization, while in the infrared (IR)
limit $q^2 \ll m^2$ they exhibit a quadratic decoupling, as
expected from the Applequist-Carazzone theorem. The decoupling
can be observed for both inverse Newton constant and for $a_3$.
With respect to the global scaling the $\,\Box R$-term is the same
as the $R^2$ term. It is well known that the finite contribution for 
the $R^2$ term is linked to the divergences of the $\,\Box R$-term, 
while the finite nonlocal contribution for the surface $\,\Box R$ 
term has smaller relevance than the one for the second derivative term.

The main new result of our work is the non-local form factors for the
Einstein-Hilbert term, which has the form $k(\Box)R$. For the
non-zero mass $m$ of the quantum field such a form factor can be
expanded into power series in the ratio $\Box/m^2$ and thus it
represents a power series of total derivatives. If we forget that the
total derivatives do not contribute to the equations of motion,
these form factors show typical quadratic decoupling in the
IR limit $q^2 \ll m^2$. The same effect can be observed from
both form factors in the effective action and from the ``physical''
beta functions defined in the Momentum Subtraction scheme
of renormalization.

The relevant question is whether there is a manner to construct
a physical application for the results for the total derivative terms.
In this respect we can note that the total derivative terms may be
relevant in the case of manifolds with boundaries. In the
theoretical cosmology there are objects of this type called domain
walls, and it would be interesting to consider the implications of
our results in this case. Even more simple is the situation in
cosmology. One can regard the cosmological spacetime of the
expanding universe as a manifold with boundary (horizon) which
has a size defined by the inverse of Hubble parameter. Taking this
into account, the natural interpretation is that we have, for the
Einstein-Hilbert term, the decoupling in the form of identification
\beq
\frac{q^2}{m^2}\,\,\longrightarrow \,\,\frac{H^2}{m^2}.
\label{idH}
\eeq
Indeed, the quadratic decoupling for the inverse Newton constant
in the IR is not what we need for the phenomenological models of
quantum corrections in cosmology \cite{CC-Gruni} or
astrophysics \cite{RotCurves}. Using the approach of  \cite{CC-Gruni}
one can easily see that in this case the energy conservation law will
tell us that the cosmological constant does not show any significant
running in the IR. This is the result which some of the present authors
could not achieve in \cite{DCCrun}. In our opinion, however, this
conclusion can not be seen as final, since it is based on the
qualitative and phenomenological identification of the scale \eqref{idH}.
Nevertheless, one can expect that the study based on surface terms can be
useful in the further exploration of this interesting subject.

\bigskip

%%%%%%%%%%%%%%%%%%%%%%%%%%%%%%%%
\noindent\emph{Acknowledgements.}
Authors are grateful to Tiago G.~Ribeiro for involvement in the
early stages of the project. O.Z.\ is grateful to Carlo~Pagani for several
discussions on this and related topics. O.Z.\ acknowledges support
from the DFG under the Grant Za~958/2-1. The work of I.Sh.\ was
partially supported by Conselho Nacional de Desenvolvimento
Cient\'{i}fico e Tecnol\'{o}gico - CNPq  (grant 303893/2014-1)
and Funda\c{c}\~{a}o de Amparo \`a Pesquisa de Minas Gerais -
FAPEMIG (project APQ-01205-16).
T.P.N.\ wishes to acknowledge
CAPES for the support through the PNPD program.
S.A.F.\ acknowledges support from  the DAAD and the Ministerio
de Educaci\'on Argentino under the ALE-ARG program.

%%%%%%%%%%%%%%%%%%%%%%
\appendix

%%%%%%%%%%%%%%%%%%%%%%%%%%%%%%%
%%%%%%%%%%%%%%%%%%%%%%%%%%%%%%%
\section{The non-local expansion of the heat kernel} \label{sect:heat-kernel}
%%%%%%%%%%%%%%%%%%%%%%%%%%%%%%%%%%%%%%%%%%%%%%%%%%%%%%%%%%%%%%%
%%%%%%%%%%%%%%%%%%%%%%%%%%%%%%%%%%%%%%%%%%%%%%%%%%%%%%%%%%%%%%%

In this Appendix we briefly present the non-local expansion of the heat kernel \cite{bavi87,bavi90,Codello:2012kq}.
Consider a Laplace-type operator
\begin{eqnarray}
 {\cal D} &=& \Delta_g + E
\end{eqnarray}
which acts on a generic tensor bundle equipped with a connection over 
a Riemaniann manifold which has Euclidean metric $g_{\mu\nu}$.
We introduced the Laplacian $\Delta_g$ which is defined as the 
negative of the square of the covariant derivative 
$\Delta_g = -\nabla^2= -g^{\mu\nu}\nabla_\mu\nabla_\nu$
and a local endomorphism $E$ which acts multiplicatively.

The (local) heat kernel ${\cal H}(s;x,x')$ is defined as the solution
of the initial value problem
\begin{eqnarray}
 (\partial_s + {\cal D}_x) {\cal H}(s;x,x') = 0\,,\qquad  {\cal H}(s;x,x') &=& \delta(x,x')\,,
\end{eqnarray}
in which $\delta(x,x')$ is the covariant Dirac delta. The heat kernel allows us to give a covariant representation
to traces of functions of the Laplace-type operator ${\cal D}$, and specifically allows us to compute the 1-loop effective action $\Ga[g]$ .
The dimensionally regularized effective action is\footnote{Notice that we use the formal notation ${\cal H} (s)=e^{-s\mathcal{D}}$
from which it follows that the heat kernel is given by the matrix values of this operator, i.e. ${\cal H} (s; x,x')=\langle x \vert \mathcal{H} \vert x'\rangle$.}
\begin{equation}
 \begin{split}
 \Ga[g] &= \frac{1}{2}\, \Tr \ln \left(\D +E +m^2\right)
 = -\frac{\mu^{\epsilon}}{2} \, \Tr \int_0^\infty \frac{{\rm d}s}{s} \, {\rm e}^{-sm^2} {\cal H}(s)\,.
 \end{split}
\end{equation}

The trace of the local heat kernel admits a curvature expansion that to the second order is
\begin{equation}
 \begin{split}
 {\Tr\, \cal H}(s) =& \frac{1}{(4\pi s)^{d/2}} \int {\rm d}^4 x \sqrt{g}\, {\rm tr} \Bigl\{
  \mathbf{1}
  + s G_E(s\Delta_g) E
  + s G_R(s\Delta_g) R\\
 &
  +s^2 R F_R(s\Delta_g)R
  +s^2 R^{\mu\nu} F_{Ric}(s\Delta_g)R_{\mu\nu}
  +s^2 E F_E(s\Delta_g)E\\
 &
  +s^2 E F_{RE}(s\Delta_g)R
  +s^2 \Omega^{\mu\nu} F_\Omega(s\Delta_g) \Omega_{\mu\nu}
 \Bigr\}
 + {\cal O}\left({\cal R}\right)^3\,,
 \end{split}
\end{equation}
in which ${\cal O}\left({\cal R}\right)^3$ represents all possible 
non-local terms with three or more curvatures \cite{bavi87,bavi90}.
The functions whose argument is $\Delta_g$ are known as form factors 
of the heat kernel: they act on the curvatures of the expansion and 
should be regarded as non-local functions of the Laplacian. The form 
factors appearing in the linear terms have been derived in 
\cite{Codello:2012kq} as
\begin{equation}
 \begin{split}
 G_E(x) = -f(x)\,,\qquad G_R(x) = \frac{f(x)}{4}+\frac{f(x)-1}{2x}\,,
 \end{split}
\end{equation}
while those appearing in the quadratic terms can be found in \cite{bavi87,bavi90,Codello:2012kq}.
All form factors depend on a single basic form factor which is defined as
\begin{equation}
 \begin{split}
 f(x) &= \int_0^1 \!{d}\alpha \, {\rm e}^{-\alpha(1-\alpha)x}\,.
 \end{split}
\end{equation}
All the form factors admit well-defined expansions both for large and 
small values of the parameter $s$, since $s$ is dual to the energy of 
the fluctuations the non-local expansion is a suitable tool to explore 
the effective action from high- to low-energies

%%%%%%%%%%%%%%%%%%%%%%%%%%%%%%%%%%%%%%%%%
%%%%%%%%%%%%%%%%%%%%%%%%%%%%%%%%%%%%%%%%%
\section{Comments on the UV structure of the effective action}\label{sect:uv-structure}
%%%%%%%%%%%%%%%%%%%%%%%%%%%%%%%%%%%%%%%%%
%%%%%%%%%%%%%%%%%%%%%%%%%%%%%%%%%%%%%%%%%

The local and non-local contributions to the effective action 
\eqref{eq:effective-action-full} are not fully independent,
but rather display some important relations which underline 
the properties described in Sect.~\ref{sect:non-local-effective-action}.
Let us concentrate here on the renormalization of a generic operator $O[g]$ 
on which a form factor $B_O(z)$ acts. (The explicit example that appears 
in the text would be to take $R$ as the operator and $B(z)$ as the 
corresponding form factor.)
For small mass $m^2\sim 0$ we notice that the regularized vacuum 
action is always of the form
\begin{equation}
 \begin{split}\label{eq:gamma-local-div}
 \Ga[g] &\supset
 - \frac{b _O}{(4\pi)^2\bar{\epsilon}}  \int {\rm d}^4x~O[g]
 + \frac{1}{2(4\pi)^2}\int {\rm d}^4x~ B_O(z)~O[g]  \\
 &= -\frac{b _O}{2(4\pi)^2}\int {\rm d}^4x \Bigl[\frac{2}{\bar{\epsilon}} - \ln\left(-\nabla^2/m^2\right)\Bigr]O[g] + \dots
 \end{split}
\end{equation}
in which the dots hide subleading contributions in the mass and $b_O$
is a pure number related to the renormalization of the operator.
The above relation underlines an explicit connection between the 
coefficient of the $1/\bar{\epsilon}$ pole and the leading ultraviolet 
logarithmic behavior of the form factor 
\cite{El-Menoufi:2015cqw,Donoghue:2015nba}.

The subtraction of the pole requires the introduction of the renormalized coupling $g_O$
\begin{equation}
 \begin{split}
 S_{\rm ren}[g] \supset \int g_O ~ O[g]\,,
 \end{split}
\end{equation}
which in the $\overline{\rm MS}$ scheme will have the beta function
\begin{equation}
 \begin{split}
 \beta^{\overline{\rm MS}}_{g_O} &= \frac{b _O}{(4\pi)^2}\,.
 \end{split}
\end{equation}
Following our discussion of Sect.~\ref{sect:non-local-effective-action} we find that
if we subtract the divergence at the momentum scale $q^2$ coming from the Fourier transform of the form factor
we get the non-local beta function
\begin{equation}\label{eq:beta-funct}
 \begin{split}
 \beta_{g_O} &= \frac{z}{(4\pi)^2} B'_{g_O}(z)\,.
 \end{split}
\end{equation}
Using \eqref{eq:gamma-local-div} it is easy to see that in the ultraviolet limit $z\gg 1$
\begin{equation}
 \begin{split}
  B(z)&= b _O \ln\left(z\right) + \dots \,,
 \end{split}
\end{equation}
from which it is easy to see in general that the ultraviolet limit of the non-local beta function coincides with the $\overline{\rm MS}$ result
\begin{equation}
 \begin{split}
  \beta_{g_O}&=\beta^{\overline{\rm MS}}_{g_O} +\dots \qquad{\rm for}\quad z\gg 1\,.
 \end{split}
\end{equation}

In the above discussion we have however always implicitly assumed that the operator $O[g]$
is kept fixed upon actions of the renormalization group operator $q\partial q=2z\partial_z$.
Suppose now that the operator $O[g]$ is actually a total derivative of the form
\begin{equation}
 \begin{split}
  O[g] = \Box \, O'[g] = -\Delta_g O'[g] \,,
 \end{split}
\end{equation}
in which we introduce another operator $O'[g]$, which itself needs to
be renormalized with a coupling $g_{O'}$ and a local term
$g_{g_{O'}}\int O'[g]$.
If we now act  with $q\partial_q$ and keep $O'[g]$ instead of $O[g]$ the renormalization
group flow will acquire an overall scaling term due to the above relation.
A solution to this problem is to manually remove such scaling and define
\begin{equation}\label{eq:beta-funct-tot-der}
 \begin{split}
  \beta_{g_{O'}} = -\frac{1}{(4\pi)^2}\, z\,
  \partial_z\left[\frac{B_O(z)}{z}\right]\,.
 \end{split}
\end{equation}
This definition ensures that the $\overline{\rm MS}$ beta function of
the coupling of the total derivative $\Box \, O'[g]$ is correctly
reproduced in the ultraviolet limit of the non-local beta function.
This is seen in the main text in the definition of $ \Psi$ \eqref{eq:def-psi}.

As a final step to this analysis we point out that there are situations in
which the definitions of \eqref{eq:beta-funct} and \eqref{eq:beta-funct-tot-der} might need to coexist because
both the operator ${\cal O}'$ and its total derivative require renormalization.
In the main text, and in particular in the definition \eqref{eq:definitions-running}
we have adopted the strategy of subtracting the leading behaviors at large and small energies
to ensure the correct scaling properties of the renormalization group running.

%%%%%%%%%%%%%%%%%%%%%%%%%%%%%%%%

\end{document}